\documentclass[final]{svjour2}
\usepackage{booktabs}
\usepackage{amsmath}
\usepackage{amssymb}
\usepackage{algorithmic}
\usepackage{algorithm}
\usepackage{graphicx}
\usepackage{url}
\usepackage[english]{babel}

\newcommand{\set}[1]{\left\{#1\right\}}
\newcommand{\iset}[1]{\mathcal{#1}}
\newcommand{\pr}[1]{\left(#1\right)}
\newcommand{\spr}[1]{\left[#1\right]}
\newcommand{\abs}[1]{{\left|#1\right|}}
\newcommand{\norm}[1]{\left\|#1\right\|}
\newcommand{\enset}[2]{\left\{#1 ,\ldots , #2\right\}}
\newcommand{\enpath}[2]{\left(#1 ,\ldots , #2\right)}
\newcommand{\vect}[1]{\spr{#1}^T}

\newcommand{\mean}[2]{\operatorname{E}_{#1}\spr{#2}}

\newcommand{\real}{\mathbb{R}}
\newcommand{\define}{\Leftarrow}
\newcommand{\funcdef}[3]{{#1}:{#2} \to {#3}}
\newcommand{\np}{\textbf{NP}}

\newcommand{\ent}[1]{\mathcal{E}\pr{#1}}
\newcommand{\rb}[1]{X_{#1}}
\newcommand{\proj}[2]{#2_{#1}}
\newcommand{\nbhd}[2]{N\pr{#1 \mid #2}}
\newcommand{\nbhdr}[3]{N_{#3}\pr{#1 \mid #2}}
\newcommand{\rank}[2]{\operatorname{rank}\pr{#1 \mid #2}}
\newcommand{\bound}[3]{\operatorname{fi}\pr{#1 \mid #2, #3}}
\newcommand{\front}[2]{\operatorname{front}\pr{#1, #2}}
\newcommand{\safe}[2]{\operatorname{safe}\pr{#1 \mid #2}}
\newcommand{\freq}[2]{#1\pr{#2}}

\begin{document}
\title{Safe Projections of Binary Data Sets}
\author{Nikolaj Tatti}
\institute{HIIT Basic Research Unit,
Laboratory of Computer and Information Science,
Helsinki University of Technology, Finland. \email{ntatti@cc.hut.fi}}
\titlerunning{Safe Projections of Binary Data Sets}
\authorrunning{Nikolaj Tatti}
\journalname{Acta Informatica}
\date{January, 2006}
\maketitle
\begin{abstract}
Selectivity estimation of a boolean query based on frequent itemsets can be solved by describing the problem by a linear program. However, the number of variables in the equations is exponential, rendering the approach tractable only for small-dimensional cases. One natural approach would be to project the data to the variables occurring in the query. This can, however, change the outcome of the linear program.
                                                                                
We introduce the concept of safe sets: projecting the data to a safe set does not change the outcome of the linear program. We characterise safe sets using graph theoretic concepts and give an algorithm for finding minimal safe sets containing given attributes. We describe a heuristic algorithm for finding almost-safe sets given a size restriction, and show empirically that these sets outperform the trivial projection.
                                                                                
We also show a connection between safe sets and Markov Random Fields and use it to further reduce the number of variables in the linear program, given some regularity assumptions on the frequent itemsets.
\end{abstract}

\section{Introduction}

Consider the following problem: given a large, sparse matrix that holds boolean values, and a boolean formula on the columns of the matrix, approximate the probability that the formula is true for a random row of the matrix. A straightforward exact solution is to evaluate the formula on each row. Now consider the same problem using instead of the original matrix a family of frequent itemsets, i.e., sets of columns where true values co-occur in a large fraction of all rows~\cite{agrawal93mining,agrawal96apriori}. An optimal solution is obtained by applying linear programming in the space of probability distributions~\cite{hailperin65inequalities,nilsson86linear,bykowski04support}, but since a distribution has exponentially many components, the number of variables in the linear program is also large and this makes the approach infeasible. However, if the target formula refers to a small subset of the columns, it may be possible to remove most of the other columns without degrading the solution; somewhat surprisingly, it is not safe to remove all columns that do not appear in the formula. In this paper we investigate the question of which columns may be safely removed. Let us clarify this scenario with the following simple example.

\begin{example}
\label{ex:project}
Assume that we have three attributes, say $a$, $b$, and $c$, and a data set $D$ having five transactions
\[
D = \set{\pr{1,0,1},\pr{0,0,1},\pr{0,1,1},\pr{1,1,0},\pr{1,0,0}}.
\]
Let us consider five itemsets, namely $a$, $b$, $c$, $ab$, and $ac$. The frequency of an itemset is the fraction of transactions in which all the attributes appearing in the itemset occur simultaneously. This gives us the frequencies $\theta_a = \frac{3}{5}$, $\theta_b = \frac{2}{5}$, $\theta_c = \frac{3}{5}$, $\theta_{ab} = \frac{1}{5}$, and $\theta_{ac} = \frac{1}{5}$. Let $\theta = \vect{\theta_a,\theta_b,\theta_c,\theta_{ab},\theta_{ac}}$. Let us now assume that we want to estimate the frequency of the formula $b\land c$. Consider now a distribution $p$ defined on these three attributes. We assume that the distribution satisfies the frequencies, that is, $p(a = 1) = \theta_a$, $p(a = 1, b = 1) = \theta_{ab}$, etc. We want to find a distribution minimising/maximising $p(b \land c = 1)$. To convert this problem into a linear program we consider $p$ as a real vector having $2^3 = 8$ elements. To guarantee that $p$ is indeed a distribution we must require that $p$ sum to $1$ and that $p \geq 0$. The requirements that $p$ must satisfy the frequencies can be expressed in a form $Ap = \theta$ for a certain $A$. In addition, $p(b \land c = 1)$ can be expressed as $c^Tp$ for a certain $c$. Thus we have transform the original problem into a linear program
\[
\min c^Tp \quad \text{ s.t. } \sum p = 1, p \geq 0, Ap = \theta.
\]
Solving this program (and also the max-version of the program) gives us an interval $I = \spr{\frac{1}{5},\frac{2}{5}}$ for possible frequencies of $p(b \land c = 1)$. This interval has the following property: A rational frequency $\eta \in I$ if and only if there is a data set having the frequencies $\theta$ and having $\eta$ as the fraction of the transactions satisfying the formula $b \land c$. If we, however, delete the attribute $a$ from the data set and evaluate the boundaries using only the frequencies $\theta_b$ and $\theta_c$, we obtain a different interval $I' = \spr{0, \frac{2}{5}}$.
\end{example}

The problem is motivated by data mining, where fast methods for computing frequent itemsets are a recurring research theme~\cite{fimi2003}. A potential new application for the problem is privacy-preserving data mining, where the data is not made available except indirectly, through some statistics. The idea of using itemsets as a surrogate for data stems from~\cite{mannila96multiple}, where inclusion-exclusion is used to approximate boolean queries. Another approach is to assume a model for the data, such as maximum entropy~\cite{pavlov03beyond}. The linear programming approach requires no model assumptions.

The boolean query scenario can be seen as a special case for the following minimisation problem: Let $K$ be the number of attributes. Given a family $\iset{F}$ of itemsets, frequencies $\theta$ for $\iset{F}$, and some function $f$ that maps any distribution defined on a set $\set{0,1}^K$ to a real number find a distribution satisfying the frequencies $\theta$ and minimising $f$. To reduce the dimension $K$ we assume that $f$ depends only on a small subset, say $B$, of items, that is, if $p$ is a distribution defined on $\set{0,1}^K$ and $\proj{B}{p}$ is $p$ marginalised to $B$, then we can write $f(p) = f(\proj{B}{p})$. The projection is done by removing all the itemsets from $\iset{F}$ that have attributes outside $B$.

The question is, then, how the projection to $B$ alters the solution of the minimisation problem. Clearly, the solution remains the same if we can always extend a distribution defined on $B$ satisfying the projected family of itemsets to a distribution defined on all items and satisfying all itemsets in $\iset{F}$. We describe sufficient and necessary conditions for this extension property. This is done in terms of a certain graph extracted from the family $\iset{F}$. We call the set $B$ safe if it satisfies the extension property.

If the set $B$ is not safe, then we can find a safe set $C$ containing $B$. We will describe an efficient polynomial-time algorithm for finding a safe set $C$ containing $B$ and having the minimal number of items. We will also show that this set is unique. We will also provide a heuristic algorithm for finding a restricted safe set $C$ having at maximum $M$ elements. This set is not necessarily a safe set and the solution to the minimisation problem may change. However, we believe that it is the best solution we can obtain using only $M$ elements.

The rest of the paper is organised as follows: Some preliminaries are described in Section~\ref{sec:prel}. The concept of a safe set is presented in Section~\ref{sec:safe} and the construction algorithm is given in Section~\ref{sec:construct}. In Section~\ref{sec:freqs} we explain in more details the boolean query scenario. In Section~\ref{sec:junction} we study the connection between safe sets and MRFs. Section~\ref{sec:restrict} is devoted to restricted safe sets. We present empirical tests in Section~\ref{sec:tests} and conclude the paper with Section~\ref{sec:concl}. Proofs for the theorems are given in Appendix.
\section{Preliminaries}
\label{sec:prel}
We begin by giving some basic definitions. A $0$--$1$ \emph{database} is a pair $\left<D,A\right>$, where $A$ is a set of items $\enset{a_1}{a_K}$ and $D$ is a \emph{data set}, that is, a multiset of subsets of $A$.

A subset $U \subseteq A$ of items is called an \emph{itemset}. We define an \emph{itemset indicator function} $\funcdef{S_U}{\set{0,1}^K}{\set{0,1}}$ such that
\[
S_U(z) = \left\{
\begin{array}{ll}
1,\qquad &  z_i = 1\text{ for all } a_i \in U \\
0,\qquad & \text{otherwise}
\end{array}
\right..
\]
Throughout the paper we will use the following notation: We denote a random binary vector of length $K$ by $\rb{} = \rb{A}$. Given an itemset $U$ we define $\rb{U}$ to be the binary vector of length $\abs{U}$ obtained from $\rb{}$ by taking only the elements corresponding to $U$.

The \emph{frequency} of the itemset $U$ taken with respect of $D$, denoted by $\freq{U}{D}$, is the mean of $S_U$ taken with respect $D$, that is, $\freq{U}{D} = \frac{1}{\abs{D}}\sum_{z \in D}S_U(z)$. For more information on itemsets, see e.g.~\cite{agrawal93mining}.

An \emph{antimonotonic family} $\iset{F}$ of itemsets is a collection of itemsets such that for each $U \in \iset{F}$ each subset of $U$ also belongs to $\iset{F}$. We define straightforwardly the itemset indicator function $S_\iset{F} = \set{S_U \mid U \in \iset{F}}$ and the frequency $\freq{\iset{F}}{D}  = \set{\freq{U}{D} \mid U \in \iset{F}}$ for families of itemsets.

If we assume that $\iset{F}$ is an ordered family, then we can treat $S_\iset{F}$ as an ordinary function $\funcdef{S_\iset{F}}{\set{0,1}^K}{\set{0,1}^L}$, where $L$ is the number of elements in $\iset{F}$. Also it makes sense to consider the frequencies $\freq{\iset{F}}{D}$ as a vector (rather than a set). We will often use $\theta$ to denote this vector. We say that a distribution $p$ defined on $\set{0,1}^K$ \emph{satisfies} the frequencies $\theta$, if $\mean{p}{S_\iset{F}} = \theta$.

Given a set of items $C$, we define a \emph{projection} operator in the following way: A data set $\proj{C}{D}$ is obtained from $D$ by deleting the attributes outside $C$. A projected family of itemsets $\proj{C}{\iset{F}} = \set{U \in \iset{F} \mid U \subseteq C}$ is obtained from $\iset{F}$ by deleting the itemsets that have attributes outside $C$. The projected frequency vector $\proj{C}{\theta}$ is defined similarly. In addition, if we are given a distribution $p$ defined on $\set{0,1}^K$, we define a distribution $\proj{C}{p}$ to be the marginalisation of $p$ to $C$. Given a distribution $q$ over $C$ we say that $p$ is an \emph{extension} of $q$ if $\proj{C}{p} = q$.

\section{Safe Projection}
\label{sec:safe}
In this section we define a safe set and describe how such sets can be characterised using certain graphs. 

We assume that we are given a set of items $A = \enset{a_1}{a_K}$ and an antimonotonic family $\iset{F}$ of itemsets and a frequency vector $\theta$ for $\iset{F}$. We define $\mathbb{P}$ to be the set of all probability distributions defined on the set $\set{0,1}^K$. We assume that we are given a function $\funcdef{f}{\mathbb{P}}{\real}$ mapping a distribution to a real number. Let us consider the following problem:
\begin{equation}
\begin{array}{lrcr}
\textsc{Problem P:} \\
\text{Minimise} & f(p) \\
\text{subject to} & p & \in & \mathbb{P} \\
& \mean{p}{S_\iset{F}} & = & \theta.
\end{array}
\label{eq:minproblem}
\end{equation}
That is, we are looking for the minimum value of $f$ among the distributions satisfying the frequencies $\theta$. Generally speaking, this is a very difficult problem. Each distribution in $\mathbb{P}$ has $2^K$ entries and for large $K$ even the evaluation of $f(p)$ may become infeasible. This forces us to make some assumptions on $f$. We assume that there is a relatively small set $C$ such that $f$ does not depend on the attributes outside $C$. In other words, we can define $f$ by a function $f_C$ such that $f_C(\proj{C}{p}) = f(p)$ for all $p$. Similarly, we define $\proj{C}{\mathbb{P}}$ to be the set of all distributions defined on the set $\set{0,1}^\abs{C}$. We will now consider the following projected problem:
\[
\begin{array}{lrcr}
\textsc{Problem P$_C$:} \\
\text{Minimise} & f_C(q) \\
\text{subject to} & q & \in & \proj{C}{\mathbb{P}} \\
& \mean{q}{S_\iset{\proj{C}{F}}} & = & \proj{C}{\theta}.
\end{array}
\]
Let us denote the minimising distribution of Problem P by $\hat{p}$ and the minimising distribution of Problem P$_C$ by $\hat{q}$. It is easy to see that $f(\hat{p}) \geq f_C(\hat{q})$. In order to guarantee that $f(\hat{p}) = f_C(\hat{q})$, we need to show that $C$ is safe as defined below.
\begin{definition}
Given an antimonotonic family $\iset{F}$ and frequencies $\theta$ for $\iset{F}$, a set $C$ is $\theta$-\emph{safe} if for any distribution $q \in \proj{C}{\mathbb{P}}$ satisfying the frequencies $\proj{C}{\theta}$, there exists an extension $p \in \mathbb{P}$ satisfying the frequencies $\theta$. If $C$ is safe for all $\theta$, we say that it is \emph{safe}.
\end{definition}
\begin{example}
\label{ex:unsafe}
Let us continue Example~\ref{ex:project}. We saw that the outcome of the linear program changes if we delete the attribute $a$. Let us now show that the set $C = \set{b,c}$ is not a safe set. Let $q$ be a distribution defined on the set $C$ such that $q(b=0,c=0) = 0$, $q(b=1,c=0) = \frac{2}{5}$, $q(b=0,c=1) = \frac{3}{5}$, and $q(b=1,c=1) = 0$. Obviously, this distribution satisfies the frequencies $\theta_b$ and $\theta_c$. However, we cannot extend this distribution to $a$ such that all the frequencies are to be satisfied. Thus, $C$ is not a safe set.
\end{example}

We will now describe a sufficient condition for safeness. We define a \emph{dependency graph} $G$ such that the vertices of $G$ are the items $V(G) = A$ and the edges correspond to the itemsets in $\iset{F}$ having two items $E(G) = \set{\set{a_i,a_j} \mid a_ia_j \in \iset{F}}$. The edges are undirected. Assume that we are given a subset $C$ of items and select $x \notin C$. A path $P = \enpath{a_{i_1}}{a_{i_L}}$ from $x$ to $C$ is a graph path such that $x =  a_{i_1}$ and only $a_{i_L} \in C$. We define a \emph{frontier} of $x$ with respect of $C$ to be the set of the last items of all paths from $x$ to $C$
\[
\front{x}{C} = \set{a_{i_L} \mid P = \enpath{a_{i_1}}{a_{i_L}} \text{ is a path from }x\text{ to } C}.
\]
Note that $\front{x}{C} = \front{y}{C}$, if $x$ and $y$ are connected by a path not going through $C$. The following theorem gives a sufficient condition for safeness.
\begin{theorem}
Let $\iset{F}$ be an antimonotonic family of itemsets. Let $C$ be a set of items $C \subseteq A$ such that for each $x \notin C$ the frontier of $x$ is in $\iset{F}$, that is, $\front{x}{C} \in \iset{F}$. It follows that $C$ is a safe set.
\label{thr:suf}
\end{theorem}
The vague intuition behind Theorem~\ref{thr:suf} is the following: $x$ has influence on $C$ only through $\front{x}{C}$. If $\front{x}{C} \in \iset{F}$, then the distributions marginalised to $\front{x}{C}$ are \emph{fixed} by the frequencies. This means that $x$ has no influence on $C$ and hence it can be removed.

We saw in Examples~\ref{ex:project}~and~\ref{ex:unsafe} that the projection changes the outcome if the projection set is not safe. This holds also in the general case:
\begin{theorem}
Let $\iset{F}$ be an antimonotonic family of itemsets. Let $C$ be a set of items $C \subseteq A$ such that there exists $x \notin C$ whose frontier is not in $\iset{F}$, that is, $\front{x}{C} \notin \iset{F}$. Then there are frequencies $\theta$ for $\iset{F}$ such that $C$ is not $\theta$-safe.
\label{thr:nec}
\end{theorem}
Safeness implies that we can extend every satisfying distribution $q$ in Problem P$_C$ to a satisfying distribution $p$ in Problem P. This implies that the optimal values of the problems are equal:
\begin{theorem}
Let $\iset{F}$ be an antimonotonic family of itemsets. If $C$ is a safe set, then the minimum value of Problem P is equal to the minimum value of Problem P$_C$ for any query function and for any frequencies $\theta$ for $\iset{F}$.
\end{theorem}
If the condition of being safe does not hold, that is, there is a distribution $q$ that cannot be extended, then we can define a query $f$ resulting $0$ if the input distribution is $q$, and $1$ otherwise. This construction proves the following theorem:
\begin{theorem}
Let $\iset{F}$ be an antimonotonic family of itemsets. If $C$ is not a safe set, then there is a function $f$ and frequencies $\theta$ for $\iset{F}$ such that the minimum value of Problem P is strictly larger than the minimum value of Problem P$_C$.
\end{theorem}
\begin{example}
\label{ex:safe}
Assume that we have $6$ attributes, namely, $\set{a,b,c,d,e,f}$, and an antimonotonic family $\iset{F}$ whose maximal itemsets are $ab$, $bc$, $cd$, $ad$, $de$, $ce$, and $af$. The dependency graph is given in Fig.~\ref{fig:ex1graph}.
\begin{figure}[ht!]
\center
\includegraphics[width=4cm]{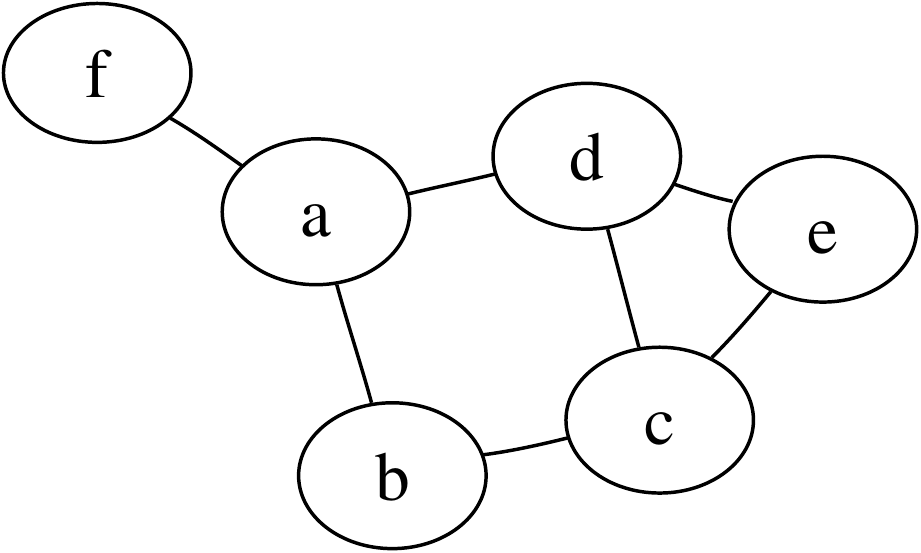}
\caption{An example of dependency graph.}
\label{fig:ex1graph}
\end{figure}

Let $C_1 = \set{a,b,c}$. This set is not a safe set since $\front{d}{C_1} = ac \notin \iset{F}$. On the other hand the set $C_2 = \set{a,b,c,d}$ is safe since $\front{f}{C_2} = a \in \iset{F}$ and $\front{e}{C_2} = cd \in \iset{F}$.
\end{example}
The proof of Theorem~\ref{thr:suf} reveals also an interesting fact:
\begin{theorem}
Let $\iset{F}$ be an antimonotonic family of itemsets and let $\theta$ be frequencies for $\iset{F}$. Let $C$ be a safe set. Let $p^{ME}$ be the maximum entropy distribution defined on $A$ and satisfying $\theta$. Let $q^{ME}$ be the maximum entropy distribution defined on $C$ and satisfying the projected frequencies $\proj{C}{\theta}$. Then $q^{ME}$ is $p^{ME}$ marginalised to $C$.
\label{thr:me}
\end{theorem}
The theorem tells us that if we want to obtain the maximum entropy distribution marginalised to $C$ and if the set $C$ is safe, then we can remove the items outside $C$. This is useful since finding maximum entropy using Iterative Fitting Procedure requires exponential amount of time~\cite{darroch72gis,jirousek95iterative}. Using maximum entropy for estimating the frequencies of itemsets has been shown to be an effective method in practice~\cite{pavlov03beyond}. In addition, if we estimate the frequencies of several boolean formulae using maximum entropy distribution marginalised to safe sets, then the frequencies are consistent. By this we mean that the frequencies are all evaluated from the same distribution, namely $p^{ME}$.

\section{Constructing a Safe Set}
\label{sec:construct}
Assume that we are given a function $f$ that depends only on a set $B$, not necessarily safe. In this section we consider a problem of finding a safe set $C$ such that $B \subseteq C$ for a given $B$. Since there are usually several safe sets that include $B$, for example, the set of all attributes $A$ is always a safe set, we want to find a safe set having the minimal number of attributes. In this section we will describe an algorithm for finding such a safe set. We will also show that this particular safe set is unique.

The idea behind the algorithm is to augment $B$ until the safeness condition is satisfied. However, the order in which we add the items into $B$ matters. Thus we need to order the items. To do this we need to define a few concepts: A \emph{neighbourhood} $\nbhd{x}{r}$ of an item $x$ of radius $r$ is the set of the items reachable from $x$ by a graph path of length at most $r$, that is,
\begin{equation}
\nbhd{x}{r} = \set{y \mid \exists P : x  \to y, \abs{P} \leq r}.
\label{eq:nbhd}
\end{equation}
In addition, we define a \emph{restricted neighbourhood} $\nbhdr{x}{y}{C}$ which is similar to $\nbhd{x}{r}$ except that now we require that only the last element of the path $P$ in Eq.~\ref{eq:nbhd} can belong to $C$. Note that $\nbhdr{x}{r}{C} \cap C \subseteq \front{x}{C}$ and that the equality holds for sufficiently large $r$.

The \emph{rank} of an item $x$ with respect of $C$, denoted by $\rank{x}{C}$, is a vector $v$ of length $\abs{A}-1$ such that $v_i$ is the number of elements in $C$ to whom the shortest path from $x$ has the length $i$, that is,
\[
v_i = \abs{C \cap \pr{\nbhdr{x}{i}{C}-\nbhdr{x}{i-1}{C}}}.
\]
We can compare ranks using the bibliographic order. In other words, if we let $v = \rank{x}{C}$ and $w = \rank{y}{C}$, then $\rank{x}{C} < \rank{y}{C}$ if and only if there is an integer $M$ such that $v_M < w_M$ and $v_i = w_i$ for all $i = 1,\ldots,M-1$.

We are now ready to describe our search algorithm. The idea is to search the items that violate the assumption in Theorem~\ref{thr:suf}. If there are several candidates, then items having the maximal rank are selected. Due to efficiency reasons, we do not look for violations by calculating $\front{x}{C}$. Instead, we check whether $\nbhdr{x}{r}{C} \cap C \in \iset{F}$. This is sufficient because
\[
\nbhdr{x}{r}{C} \cap C \notin \iset{F} \implies \front{x}{C} \notin \iset{F}.
\]
This is true because $\nbhdr{x}{r}{C} \cap C \subseteq \front{x}{C}$ and $\iset{F}$ is antimonotonic.
The process is described in full detail in Algorithm~\ref{cd:search}.
\begin{algorithm}[ht!]
\caption{The algorithm for finding a safe set $C$. The required input is $B$, the set that should be contained in $C$, and an antimonotonic family $\iset{F}$ of itemsets. The graph $G$ is the dependency graph evaluated from $\iset{F}$.}
\begin{algorithmic}
\STATE $C \define B$.
\REPEAT
\STATE $r \define 1$.
\STATE $V \define \set{x \mid \exists y \in C, xy \in E(G)}-C$ \COMMENT{$V$ contains the neighbours of $C$.}
\REPEAT
\STATE For each $x \in V$, $U_x \define \nbhdr{x}{r}{C} \cap C$.
\IF{there exists $U_x$ such that $U_x \notin \iset{F}$}
\STATE \textbf{Break} \COMMENT{A violation is found.}
\ENDIF
\STATE $r \define r+1$.
\UNTIL{no $U_x$ changed}
\IF{there is a violation}
\STATE $W \define \set{x \in V \mid U_x \notin \iset{F}}$ \COMMENT{$W$ contains the violating items.}
\STATE $v \define \max\set{\rank{x}{C} \mid x \in W}$.
\STATE $Z \define \set{x \in W \mid \rank{x}{C} = v}$
\STATE $C \define C \cup Z$ \COMMENT{Augment $C$ with the violating items having the largest rank.}
\ENDIF
\UNTIL{there are no violations.}
\end{algorithmic}
\label{cd:search}
\end{algorithm}

We will refer to the safe set Algorithm~\ref{cd:search} produces as $\safe{B}{\iset{F}}$. We will now show that $\safe{B}{\iset{F}}$ is the smallest possible, that is,
\[
\abs{\safe{B}{\iset{F}}} = \min \set{\abs{Y} \mid B \subseteq Y, Y \text{ is a safe set}}.
\]
The following theorem shows that in Algorithm~\ref{cd:search} we add only necessary items into $C$ during each iteration.
\begin{theorem}
Let $C$ be a set of items during some iteration of Algorithm~\ref{cd:search} and let $Z = \set{x \in W \mid \rank{x}{C} = v}$ be the set of items as it is defined in Algorithm~\ref{cd:search}. Let $Y$ be any safe set containing $C$. Then it follows that $Z \subseteq Y$.
\label{thr:optimal}
\end{theorem}
\begin{corollary}
\label{cor:unique}
A safe set containing $B$ containing the minimal number of items is unique. Also, this set is contained in each safe set containing $B$.
\end{corollary}
\begin{corollary}
Algorithm~\ref{cd:search} produces the optimal safe set.
\end{corollary}
\begin{example}
Let us continue Example~\ref{ex:safe}. Assume that our initial set $B$ is $\set{a, b, c}$. We note that $\front{d}{B} = \front{e}{B} = ac \notin \iset{F}$. Therefore, $B$ is not a safe set. The ranks are $\rank{d}{B} = 2$ and $\rank{e}{B} = \vect{1, 1}$ (the trailing zeros are removed). It follows that the rank of $d$ is larger than the rank of $e$ and therefore $d$ is added into $B$ during Algorithm~\ref{cd:search}. The resulting set $C = \set{a, b, c, d}$ is the minimal safe set containing $B$.
\end{example}
\section{Frequencies of Boolean Formulae.}
\label{sec:freqs}
A boolean formula $\funcdef{f}{\set{0,1}^K}{\set{0,1}}$ maps a binary vector to a binary value. Given a family $\iset{F}$ of itemsets and frequencies $\theta$ for $\iset{F}$ we define a \emph{frequency interval}, denoted by $\bound{f}{\iset{F}}{\theta}$, to be
\[
\bound{f}{\iset{F}}{\theta} = \set{\mean{p}{f} \mid \mean{p}{S_\iset{F}} = \theta},
\]
that is, a set of possible frequencies coming from the distribution satisfying given frequencies. For example, if the formula $f$ is of form $a_1 \land \ldots \land a_M$, then we are approximating the frequency of a possibly unknown itemset.

Note that this set is truly an interval and its boundaries can be found using the optimisation problem given in Eq.~\ref{eq:minproblem}. It has been shown that finding the boundaries can be reduced to a linear programming~\cite{hailperin65inequalities,nilsson86linear,bykowski04support}. However, the problem is exponential in $K$ and therefore it is crucial to reduce the dimension. Let us assume that the boolean formula depends only on the variables coming from some set, say $B$. We can now use Algorithm~\ref{cd:search} to find a safe set $C$ including $B$ and thus to reduce the dimension.
\begin{example}
\label{ex:freqs}
Let us continue Example~\ref{ex:safe}. We assign the following frequencies to the itemsets: $\theta_x = 0.5$ where $x \in \set{a,b,c,d,e,f}$, $\theta_{bd} = 0.5$, $\theta_{cd} = 0.4$, and the frequencies of the rest itemsets in $\iset{F}$ are equal to $0.25$. We consider the formula $f = b \land c$. In this case $f$ depends only on $B = \set{b,c}$. If we project directly to $B$, then the frequency is equal to $\bound{f}{\proj{B}{\iset{F}}}{\proj{B}{\theta}} = \spr{0,0.5}$.

The minimal safe set containing $B$ is $C = \set{a,b,c,d}$. Since $\theta_{bd} = 0.5$ it follows that $b$ is equivalent to $d$. This implies that the frequency of $f$ must be equal to $\bound{f}{\proj{C}{\iset{F}}}{\proj{C}{\theta}} = \theta_{cd} = 0.4$.
\end{example}

There exists many problems similar to ours: A well-studied problem is called \textsc{PSAT} in which we are given a CNF-formula and probabilities for each clause asking whether there is a distribution satisfying these probabilities. This problem is \np-complete~\cite{georgakopoulos88psat}. A reduction technique for the minimisation problem where the constraints and the query are allowed to be conditional is given in~\cite{lukasiewicz97deduction}. However, this technique will not work in our case since we are working only with unconditional queries. A general problem where we are allowed to have first-order logic conditional sentences as the constraints/queries is studied in~\cite{lukasiewicz01logic}. This problem is shown to be \np-complete. Though these problems are of more general form they can be emulated with itemsets~\cite{calders04computational}. However, we should note that in the general case this construction does not result an antimonotonic family.

There are many alternative ways of approximating boolean queries based on statistics: For example, the use of wavelets has been investigated in~\cite{matias98wavelet}. Query estimation using histograms was studied in~\cite{muralikrishna88histogram} (though this approach does not work for binary data). We can also consider assigning some probability model to data such as Chow-Liu tree model or mixture model (see e.g.~\cite{pavlov01query,pavlov03beyond,chow68treemodel}). Finally, if $B$ is an itemset and we know all the proper subsets of $B$ and $B$ is safe, then to estimate the frequency of $B$ we can use inclusion-exclusion formulae given in~\cite{calders02mining}.

\section{Safe Sets and Junction Trees}
\label{sec:junction}
Theorem~\ref{thr:suf} suggests that there is a connection between safe sets and Markov Random Fields (see e.g.~\cite{jordan99graphical} for more information on MRF). In this section we will describe how the minimal safe sets can be obtained from junction trees. We will demonstrate through a counter-example that this connection cannot be used directly. We will also show that we can use junction trees to reformulate the optimisation problem and possibly reduce the computational burden.

\subsection{Safe Sets and Separators}
Let us assume that the dependency graph $G$ obtained from a family $\iset{F}$ of itemsets is \emph{triangulated}, that is, the graph does not contain chordless circuits of size $4$ or larger. In this case we say that $\iset{F}$ is triangulated. For simplicity, we assume that the dependency graph is connected. We need some concepts from Markov Random Field theory (see e.g.~\cite{jordan99graphical}): The \emph{clique graph} is a graph having cliques of $G$ as vertices and two vertices are connected if the corresponding cliques share a mutual item. Note that this graph is connected. A spanning tree of the clique graph is called a \emph{junction tree} if it has a \emph{running intersection} property. By this we mean that if two cliques contain the same item, then each clique along the path in the junction tree also contains the same item. An edge between two cliques is called a \emph{separator}, and we associate with each separator the set of items mutual to both cliques.

We also make some further assumptions concerning the family $\iset{F}$: Let $V$ be the set of items of some clique of the dependency graph. We assume that every proper subset of $V$ is in $\iset{F}$. If $\iset{F}$ satisfies this property for each clique, then we say that $\iset{F}$ is \emph{clique-safe}. We do not need to have $V \in \iset{F}$ because there is no node having an entire clique as a frontier. 

Let us now investigate how safe sets and junction trees are connected. First, fix some junction tree, say $T$, obtained from $G$. Assume that we are given a set $B$ of items, not necessarily safe. For each item $b \in B$ we select some clique $Q_b \in V(T)$ such that $b \in Q_b$ (same clique can be associated with several items). Let $b, c \in B$ and consider the path in $T$ from $Q_b$ to $Q_c$. We call the separators along such paths \emph{inner separators}. The other separators are called \emph{outer separators}. We always choose cliques $Q_b$ such that the number of inner separators is the smallest possible. This does not necessarily make the choice of the cliques unique, but the set of inner separators is always unique. We also define an \emph{inner clique} to be a clique incident to some inner separator. We refer to the other cliques as \emph{outer cliques}.

\begin{example}
\label{ex:junction}
Let us assume that we have $5$ items, namely $\set{a,b,c,d,e}$. The dependency graph, its clique graph, and the possible junction trees are given in Figure~\ref{fig:trees}.
\begin{figure}[ht!]
\center
\begin{minipage}{3.5cm}
\includegraphics[width=3cm]{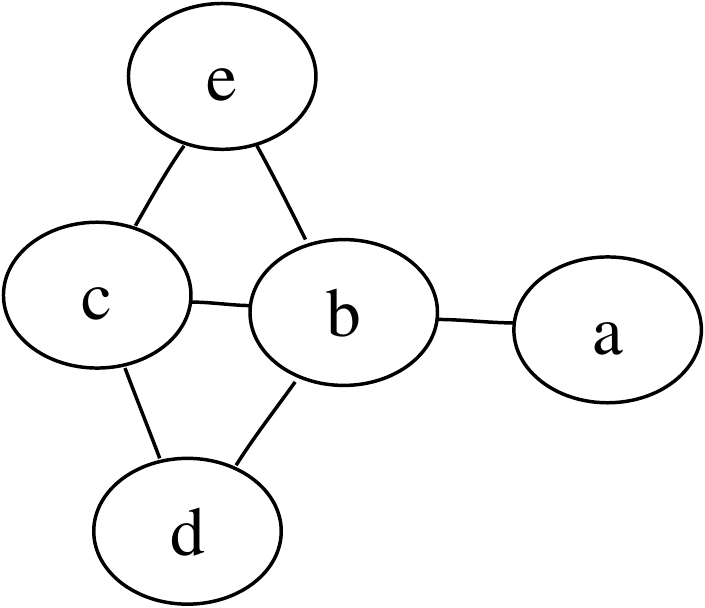}
\end{minipage}
\begin{minipage}{2.5cm}
\includegraphics[width=2cm]{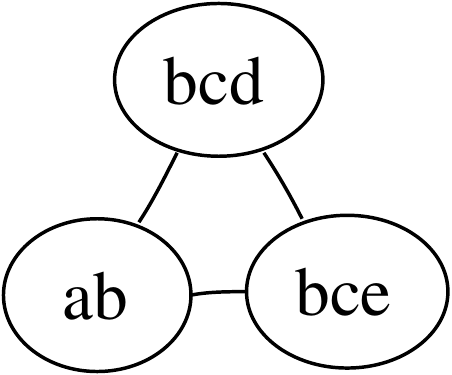}
\end{minipage}
\begin{minipage}{3.5cm}
\includegraphics[width=3cm]{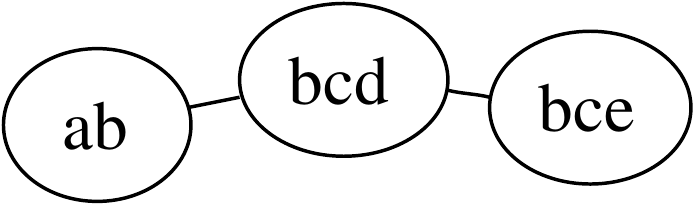}
\includegraphics[width=3cm]{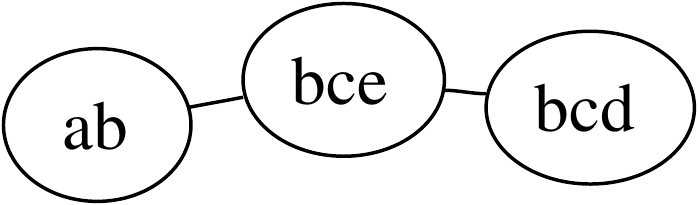}
\end{minipage}
\caption{An example of an dependency graph, a corresponding clique graph, and the possible junction trees.}
\label{fig:trees}
\end{figure}

Let $B = \set{a, d}$. Then the inner separator in the upper junction tree is the left edge. In the lower junction tree both edges are inner separators.
\end{example}

The following three theorems describe the relation between the safe sets containing $B$ and the inner separators.
\begin{theorem}
Let $\iset{F}$ be an antimonotonic, triangulated and clique-safe family of itemsets. Let $T$ be a junction tree. Let $C$ be a set containing $B$ and all the items from the inner separators of $B$. Then $C$ is a safe set.
\label{thr:junction1}
\end{theorem}
The following corollary follows from Corollary~\ref{cor:unique}.
\begin{corollary}
Let $\iset{F}$ be an antimonotonic, triangulated and clique-safe family of itemsets. Let $T$ be a junction tree. The minimal safe set containing $B$ may contain (in addition to the set $B$) only items from the inner separators of $B$.
\label{thr:junction2}
\end{corollary}

\begin{theorem}
\label{thr:junction3}
Let $\iset{F}$ be an antimonotonic, triangulated and clique-safe family of itemsets. There exists a junction tree such that the minimal safe set is precisely the set $B$ and the items from the inner separators of $B$.
\end{theorem}
Theorem~\ref{thr:junction3} raises the following question: Is there a tree, \emph{not} depending on $B$, such that the minimal safe set is precisely the set $B$ and the items from the inner separators. Unfortunately, this is not the case as the following example shows.
\begin{example}
Let us continue Example~\ref{ex:junction}. Let $B_1 = \set{a, d}$ and $B_2 = \set{a, e}$. The corresponding minimal safe sets are $C_1 = \set{a, b, d}$ and $C_2 = \set{a, b, e}$. The first case corresponds to the upper junction tree given in Figure~\ref{fig:trees}, and the latter case corresponds the lower junction tree.
\end{example}
\subsection{Reformulation of the Optimisation Problem Using Junction Trees}
\label{sec:optjunction}
We have seen that a optimisation problem can be reduced to a problem having $2^\abs{C}$ variables, where $C$ is a safe set. However, it may be the case that $C$ is very large. For example, imagine that the dependency graph is a single path $\enpath{a_{i_1}}{a_{i_L}}$ and we are interested in finding the frequency for $a_{i_1} \land a_{i_L}$. Then the safe set contains the entire path. In this section we will try to reduce the computational burden even further.

The main benefit of MRF is that we are able to represent the distribution as a fraction of certain distributions. We can use this factorisation to encode the constraints. A small drawback is that we may not be able to express easily the distribution defined on $B$, the set of which the query depends. This happens when $B$ is not contained in any clique. This can be remedied by adding edges to the dependency graph.

Let us make the previous discussion more rigorous. Let $f$ be a query function and let $B$ be the set of attributes of which $f$ depends. Let $C = \safe{B}{\iset{F}}$ be the minimal safe set containing $B$. Project the items outside $C$ and let $G$ be the connectivity graph obtained from $\proj{C}{\iset{F}}$. We add some additional edges to $G$. First, we make the set $B$ fully connected. Second, we triangulate the graph. Let $T$ be a junction tree of the resulting graph.

Since $B$ is fully connected, there is a clique $Q_r$ such that $B \subseteq Q_r$. For each clique $Q_i$ in $T$ we define $p_i$ to be a distribution defined on $Q_i$. Similarly, for each separator $S_j$ we define $q_j$ to be a distribution defined on $S_j$. Denote by~$\mathbb{S}_i$ the collection of separators of a clique $Q_i$.

\begin{equation}
\begin{array}{ll}
\textsc{Problem LP:} \\
\text{Minimise} & f(p_r) \\
\text{subject to} & \text{For each }Q_i \in V(T), \\
                  & \quad p_i \text{ satisfies } \proj{Q_i}{\theta}\\
                  & \quad p_i  \text{ is an extension of } q_j \\
                  & \quad\text{for each } S_j \in \mathbb{S}_i.
\end{array}
\label{eq:fast}
\end{equation}
The following theorem states that the above formulation is correct:
\begin{theorem}
The problem in Eq.~\ref{eq:fast} solves correctly the optimisation problem.
\label{thr:fast}
\end{theorem}
Note that we can remove all $q_j$ by combining the constraining equations. Thus we have replaced the original optimisation problem having $2^\abs{C}$ variables with a problem having $\sum 2^{\abs{Q_i}}$ variables. The number of cliques in $T$ is bounded by $\abs{C}$, the number of attributes in the safe set. To see this select any leaf clique $Q_i$. This clique must contain a variable that is not contained in any other clique because otherwise $Q_i$ is contained in its parent clique. We remove $Q_i$ and repeat this procedure. Since there are only $\abs{C}$ attributes, there can be only $\abs{C}$ cliques. Let $M$ be the size of the maximal clique. Then the number of variables is bounded by $\abs{C}2^{M}$. If $M$ is small, then solving the problem is much easier than the original formulation.
\begin{example}
Assume that we have a family of itemsets whose dependency graph $G$ is a path $\enpath{a_{i_1}}{a_{i_L}}$ and that we want to evaluate the boundaries for a formula $a_{i_1} \land a_{i_L}$. We cannot neglect any variable inside the path, hence we have a linear program having $2^L$ variables.

By adding the edge $\set{a_{i_1}, a_{i_L}}$ to $G$ we obtain a cycle. To triangulate the graph we add the edges $\set{a_{i_1}, a_{i_j}}$ for $3 \leq j \leq L-1$.
The junction tree in consists of $L-2$ cliques of the form $a_{i_1}a_{i_j}a_{i_{j+1}}$, where $2 \leq j \leq L-1$. The reformulation of the linear program gives us a program containing only $\pr{L-2}2^3$ variables.
\end{example}

\section{Restricted Safe Sets}
\label{sec:restrict}
Given a set $B$ Algorithm~\ref{cd:search} constructs the minimal safe set $C$. However, the set $C$ may still be too large. In this section we will study a scenario where we require that the set $C$ should have $M$ items, at maximum. Even if such a safe set may not exist we will try to construct $C$ such that the solution of the original minimisation problem described in Eq.~\ref{eq:minproblem} does not alter. As a solution we will describe a heuristic algorithm that uses the information available from the frequencies.

First, let us note that in the definition of a safe set we require that we can extend the distribution for any frequencies. In other words, we assume that the frequencies are the worst possible. This is also seen in Algorithm~\ref{cd:search} since the algorithm does not use any information available from the frequencies.

Let us now consider how we can use the frequencies. Assume that we are given a family $\iset{F}$ of itemsets and frequencies $\theta$ for $\iset{F}$. Let $C$ be some (not necessarily a safe) set. Let $x \notin C$ be some item violating the safeness condition. Assume that each path from $x$ to $C$ has an edge $e = (u,v)$ having the following property: Let $\theta_{uv}$, $\theta_u$, and $\theta_v$ be the frequencies of the itemsets $uv$, $u$, and $v$, respectively. We assume that $\theta_{uv} = \theta_u\theta_v$ and that the itemset $uv$ is not contained in any larger itemset in $\iset{F}$. We denote the set of such edges by $E$.

Let $W$ be the set of items reachable from $x$ by paths not using the edges in $E$. Note that the set $W$ has the same property than $x$. We argue that we can remove the set $W$. This is true since if we are given a distribution $p$ defined on $A-W$, then we can extend this distribution, for example, by setting $p(\rb{A}) = p^{ME}(\rb{W})p(\rb{A-W})$, where $p^{ME}(\rb{W})$ is the maximum entropy distribution defined on $W$. Note that if we remove the edges $E$, then Algorithm~\ref{cd:search} will not include $W$.

Let us now consider how we can use this situation in practice. Assume that we are given a function $w$ which assign to each edge a non-negative weight. This weight represents the correlation of the edge and should be $0$ if the independence assumption holds. Assume that we are given an item $x \notin C$ violating the safeness condition but we cannot afford adding $x$ into $C$. Define $H$ to be the subgraph containing $x$, the frontier $\front{x}{C}$ and all the intermediate nodes along the paths from $x$ to $C$. We consider finding a set of edges $E$ that would cut $x$ from its frontier and have the minimal cost $\sum_{e \in E} w(e)$. This is a well-known min-cut problem and it can be solved efficiently (see e.g.~\cite{papadimitriou98combi}). We can now use this in our algorithm in the following way: We build the minimal safe set containing the set $B$. For each added item we construct a cut with a minimal cost. If the safe set is larger than a constant $M$, we select from the cuts the one having the smallest weight. During this selection we neglect the items that were added before the constraint $M$ was exceeded. We remove the edges and the corresponding itemsets and restart the construction. The algorithm is given in full detail in Algorithm~\ref{cd:ressearch}.

\begin{algorithm}[ht!]
\caption{The algorithm for finding a restricted safe set $C$. The required input is $B$, the set that should be contained in $C$, an antimonotonic family $\iset{F}$ of itemsets, a constant $M$ which is an upper bound for $\abs{C}$, and a weight function $w$ for the edges. The graph $G$ is the dependency graph evaluated from $\iset{F}$.}
\begin{algorithmic}
\STATE $C \define B$.
\REPEAT
\STATE Find a violating item $x$ having the largest rank.
\IF{$\abs{C}+1 > M$}
\STATE Let $H$ be the graph containing $x$, $\front{x}{C}$ and all the intermediate nodes.
\STATE Let $E_x$ be the min-cut of $H$ cutting $x$ and $\front{x}{C}$ from each other.
\STATE Let $v_x$ be the cost of $E_x$.
\ENDIF
\STATE $C \define C + x$.
\UNTIL{there are no violations.}
\IF{$\abs{C} > M$}
\STATE Let $x$ be the item such that $v_x$ is the smallest possible.
\STATE Remove the edges $E_x$ from the dependency graph.
\STATE Remove the itemsets corresponding to the edges from $\iset{F}$.
\STATE Remove also possible higher-order itemsets to preserve the antimonotonicity of $\iset{F}$.
\STATE Restart the algorithm.
\ENDIF
\end{algorithmic}
\label{cd:ressearch}
\end{algorithm}
\begin{example}
We continue Example~\ref{ex:freqs}. As a weight function for the edges we use the mutual information. This gives us $w_{bd} = 0.6931$ and $w_{cd} = 0.1927$. The rest of the weights are $0$. Let $B = \set{b, c}$. We set the upper bound for the size of the safe set to be $M = 3$. The minimal safe set is $C = \set{a,b,c,d}$. The min cuts are $E_a = \set{\pr{a,b},\pr{a,c}}$ and $E_d = \set{\pr{d,b},\pr{d,c}}$. The corresponding weights are $v_a = 0$ and $v_d = w_{bd}+w_{cd} > 0$. Thus by cutting the edges $E_a$ we obtain the set $C^r = \set{b,c,d}$. The frequency interval for the formula $b \land c$ is $\bound{f}{\proj{C^r}{\iset{F}}}{\proj{C^r}{\theta}} = 0.4$ which is the same as in Example~\ref{ex:freqs}.
\end{example}

\section{Empirical Tests}
\label{sec:tests}
We performed empirical tests to assess the practical relevance of the restricted safe sets, comparing it to the (possibly) unsafe trivial projection. We mined itemset families from two data sets, and estimated boolean queries using both the safe projection and the trivial projection. The first data set, which we call \emph{Paleo}\footnote{\emph{Paleo} was constructed from NOW public release 030717 available from~\cite{forselius05now}.}, describes fossil findings: the attributes correspond to genera of mammals, the transactions to excavation sites. The \emph{Paleo} data is sparse, and the genera and sites exhibit strong correlations. The second data set, which we call \emph{Mushroom}, was obtained from the FIMI repository\footnote{\url{http://fimi.cs.helsinki.fi}}. The data is relatively dense.

First we used the \textsc{Apriori}~\cite{agrawal96apriori} algorithm to retrieve some families of itemsets. A problem with \textsc{Apriori} was that the obtained itemsets were concentrated on the attributes having high frequency. A random query conducted on such a family will be safe with high probability --- such a query is trivial to solve. More interesting families would the ones having almost all variables interacting with each other, that is, their dependency graphs have only a small number of isolated nodes. Hence we modified \textsc{APriori}: Let $A$ be the set containing all items and for each $a \in A$ let $m(a)$ be the frequency of $a$. Let $m$ be the smallest frequency $m = \min_{a \in A} m(a)$ and define $s(a) = m(a)/m$. Let $U$ be an itemset and let $\theta_U$ be its frequency. Define $\eta_U = \prod_{a \in U} s(a)$. We modify \textsc{Apriori} such that the itemset $U$ is in the output if and only if the ratio $\theta_U/\eta_U$ is larger than given threshold $\sigma$. Note that this family is antimonotonic and so \textsc{Apriori} can be used. By this modification we are trying to give sparse items a fair chance and in our tests the relative frequencies did produce more scattered families.

\begin{figure}[ht!]
\center
\small
\begin{minipage}{5cm}
\center
\includegraphics[width=5cm]{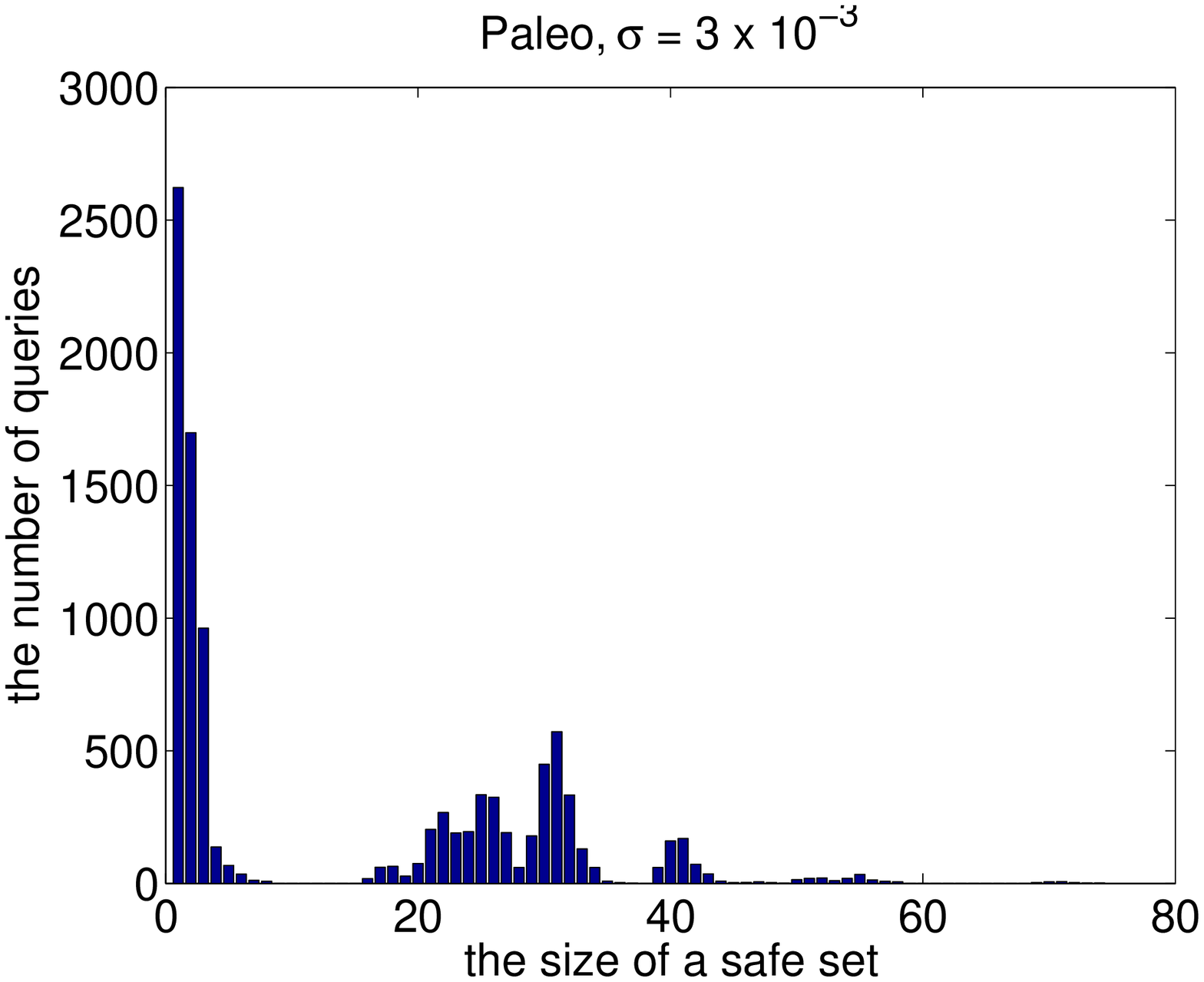}
\end{minipage}
\begin{minipage}{5cm}
\center
\includegraphics[width=5cm]{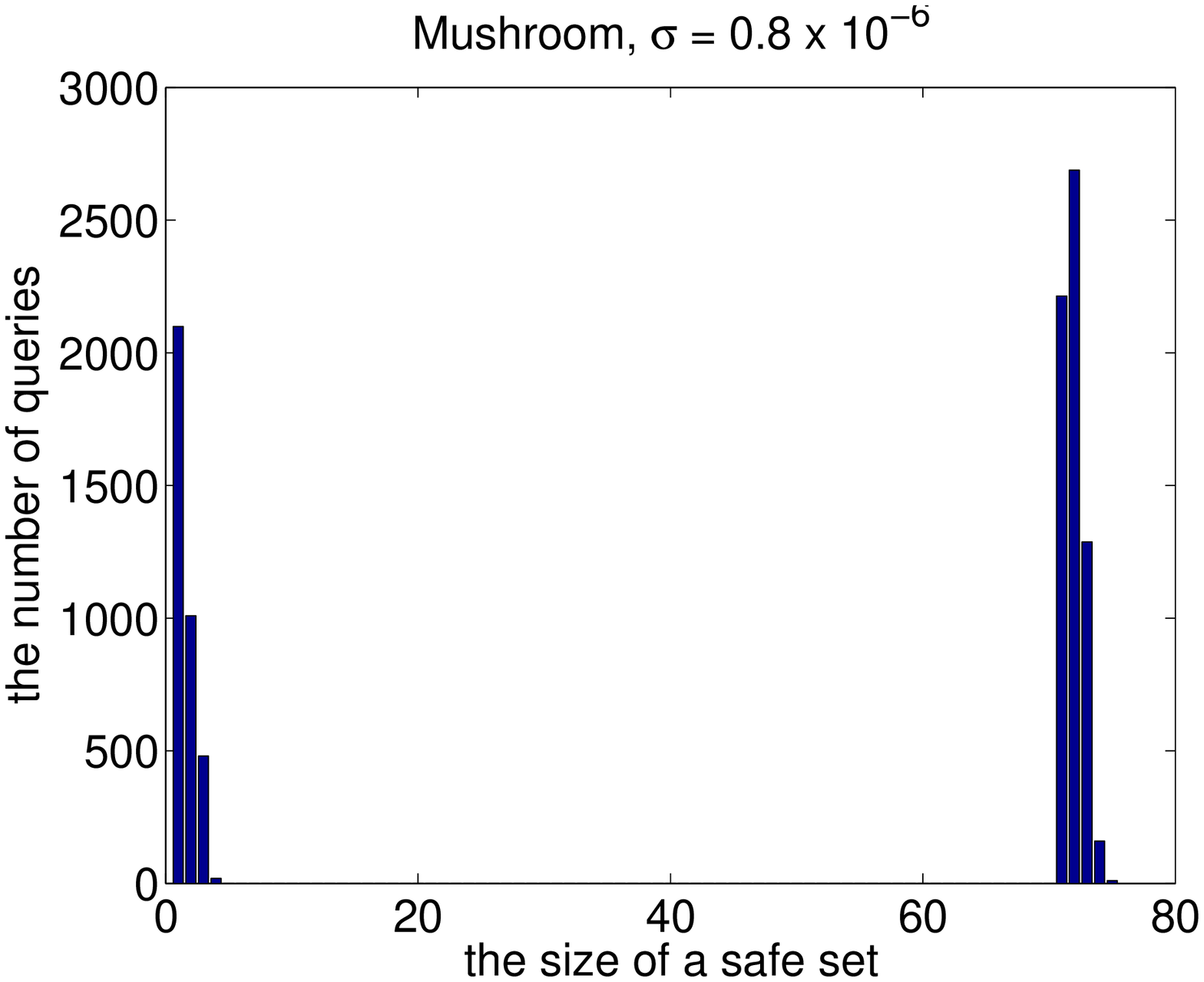}
\end{minipage}
\caption{Distributions of the sizes of safe sets. The left histogram is obtained from \emph{Paleo} data by using $\sigma = 3\times 10^{-3}$ as the threshold parameter for modified \textsc{APriori}. The right histogram is obtained from \emph{Mushroom} data with $\sigma = 0.8\times 10^{-8}$.}
\label{fig:safehist}
\end{figure}

For each family of itemsets we evaluated $10000$ random boolean queries. We varied the size of the queries between $2$ and $4$. At first, such queries seem too simple but our initial experiments showed that these queries do result large safe sets. A few examples are given in Figure~\ref{fig:safehist}. In most of the queries the trivial projection is safe but there are also very large safe sets. Needless to say that we are forced to use restricted safe sets.

Given a query $f$ we calculated two intervals $i_1(f) = \bound{f}{\proj{B}{\iset{F}}}{\proj{B}{\theta}}$ and $i_2(f) = \bound{f}{\proj{C}{\iset{F}}}{\proj{C}{\theta}}$ where $B$ contains the attributes of $f$ and $C$ is the restricted safe set obtained from $B$ using Algorithm~\ref{cd:ressearch}. In other words, $i_1(f)$ is obtained by using the trivial projection and $i_2(f)$ is obtained by projecting to the restricted safe set. As parameters for Algorithm~\ref{cd:ressearch} we set the upper bound $M = 8$ and the weight function $w$ to be the mutual information.

We divided queries into two classes. A class \textsc{Trivial} contained the queries in which the trivial projection and the restricted safe set were equal. The rest of the queries were labelled as \textsc{Complex}. We also defined a class \textsc{All} that contained all the queries.

As a measure of goodness for a frequency interval we considered the difference between the upper and the lower bound. Clearly $i_2(f) \subseteq i_1(f)$, so if we define a ratio $r(f) = \frac{\norm{i_2(f)}}{\norm{i_1(f)}}$, then it is always guaranteed that $0 \leq r(f) \leq 1$. Note that the ratio for the queries in \textsc{Trivial} is always $1$.

The ratios were divided into appropriate bins. The results obtained from \emph{Paleo} data are shown in the contingency table given in Tables~\ref{tab:paleo1}~and~\ref{tab:paleo2} and the results for \emph{Mushroom} data are given in Tables~\ref{tab:mushroom1}~and~\ref{tab:mushroom2}.

\begin{table}[ht!]
\centering
\begin{tabular}{rrrrrrrr}
\toprule
&&& \multicolumn{5}{c}{$\sigma\times10^{-3}$}\\
\cmidrule(l){4-8}

Class & $r\ge$ & $r<$ & $3$ & $3.25$ & $3.5$ & $3.75$ & $4$ \\
\midrule
\textsc{Complex}
& $0$ & $0.2$ & $1$ & $0$ & $0$ & $0$ & $0$ \\
& $0.2$ & $0.4$ & $0$ & $1$ & $1$ & $0$ & $0$ \\
& $0.4$ & $0.6$ & $15$ & $11$ & $10$ & $5$ & $4$ \\
& $0.6$ & $0.8$ & $74$ & $53$ & $50$ & $55$ & $45$ \\
& $0.8$ & $1$ & $238$ & $173$ & $124$ & $99$ & $68$ \\
& $1$ & $$ & $3289$ & $1931$ & $1353$ & $1116$ & $868$ \\
\textsc{Trivial}
& $1$ & $$ & $6383$ & $7831$ & $8462$ & $8725$ & $9015$ \\
\bottomrule
\end{tabular}
\caption{Counts of queries obtained from \emph{Paleo} data and classified according to the ratio $r(f)$, giving the relative tightness of the bounds from restricted safe sets compared to the trivial projections. A column represents a family of itemsets used as the constraints. The parameter $\sigma$ is the threshold given to the modified \textsc{APriori}. The class \textsc{Trivial} contains the queries in which the projections were equal; \textsc{Complex} contains the remaining queries. For example, there were $15$ complex queries having the ratios between $0.4-0.6$ in the first family.}
\label{tab:paleo1}
\end{table}

\begin{table}[htbp]
\centering
\begin{tabular}{rrrrrr}
\toprule
& \multicolumn{5}{c}{$\sigma\times10^{-3}$}\\
\cmidrule(l){2-6}

Class & $3$ & $3.25$ & $3.5$ & $3.75$ & $4$ \\
\midrule
\textsc{Complex} & $91.0\%$ & $89.0\%$ & $88.0\%$ & $87.5\%$ & $88.1\%$ \\
\textsc{All}     & $96.7\%$ & $97.6\%$ & $98.1\%$ & $98.4\%$ & $98.8\%$ \\
\bottomrule
\end{tabular}
\caption{Probability of $r(f)=1$ among the complex queries and among all queries. The queries were obtained from \emph{Paleo} data. A column represents a family of itemsets used as the constraints. The parameter $\sigma$ is the threshold given to the modified \textsc{APriori}.}

\label{tab:paleo2}
\end{table}

\begin{table}[ht!]
\centering
\begin{tabular}{rrrrrr}
\toprule
&&& \multicolumn{3}{c}{$\sigma\times10^{-6}$}\\
\cmidrule(l){4-6}
Class & $r\ge$ & $r<$ & $0.8$ & $0.9$ & $1$ \\
\midrule

\textsc{Complex}
& $0.0$ & $0.2$ & $46$ & $38$ & $42$ \\
& $0.2$ & $0.4$ & $96$ & $81$ & $80$ \\
& $0.4$ & $0.6$ & $302$ & $261$ & $260$ \\
& $0.6$ & $0.8$ & $96$ & $86$ & $69$ \\
& $0.8$ & $1$ & $168$ & $118$ & $109$ \\
& $1$ & & $4738$ & $4146$ & $3993$ \\
\textsc{Trivial}
& $1$ & & $4554$ & $5270$ & $5447$ \\
\bottomrule
\end{tabular}
\caption{Counts of queries obtained from \emph{Mushroom} data and classified according to the ratio $r(f)$, giving the relative tightness of the bounds from restricted safe sets compared to the trivial projections. A column represents a family of itemsets used as the constraints. The parameter $\sigma$ is the threshold given to the modified \textsc{APriori}. The class \textsc{Trivial} contains the queries in which the projections were equal; \textsc{Complex} contains the remaining queries.}
\label{tab:mushroom1}
\end{table}

\begin{table}[htbp]
\centering
\begin{tabular}{rrrr}
\toprule
& \multicolumn{3}{c}{$\sigma\times10^{-6}$}\\
\cmidrule(l){2-4}
Class & $0.8$ & $0.9$ & $1$ \\
\midrule
\textsc{Complex} & $87.0\%$ & $87.7\%$ & $87.7\%$ \\
\textsc{All}     & $92.9\%$ & $94.2\%$ & $94.4\%$ \\
\bottomrule
\end{tabular}
\caption{Probability of $r(f)=1$ among the complex queries and among all queries. The queries were obtained from \emph{Mushroom} data. A column represents a family of itemsets used as the constraints. The parameter $\sigma$ is the threshold given to the modified \textsc{APriori}.}
\label{tab:mushroom2}
\end{table}

By examining Tables~\ref{tab:paleo1}~and~\ref{tab:paleo2} we conclude the following: If we conduct a random query of form $f$, then in $97\%-99\%$ of the cases the frequency intervals are equal $i_1(f) = i_2(f)$. However, if we limit ourselves to the cases where the projections differ (the class \textsc{Complex}), then the frequency interval is equal only in about $90\%$ of the cases. In addition, the probability of $i_1(f)$ being equal to $i_2(f)$ increases as the threshold $\sigma$ grows.

The same observations apply to the results for \emph{Mushroom} data (Tables~\ref{tab:mushroom1}~and~\ref{tab:mushroom2}): In $93\%-94\%$ of the cases the frequency intervals are equal $i_1(f) = i_2(f)$, but if we consider only the cases where projections differ, then the percentage drops to $88\%$. The percentages are slightly smaller than those obtained from \emph{Paleo} data and also there are relatively many queries whose ratios are very small.

The computational burden of a trivial query is equivalent for both trivial projection and restricted safe set. Hence, we examine complex queries in which there is an actual difference in the computational burden. The results suggest that in abt. $10\%$ of the complex queries the restricted safe sets produced tighter interval.

\section{Conclusions}
We started our study by considering the following problem: Given a family $\iset{F}$ of itemsets, frequencies for $\iset{F}$, and a boolean formula find the bounds of the frequency of the formula. This can be solved by linear programming but the problem is that the program has an exponential number of variables. This can be remedied by neglecting the variables not occurring in the boolean formula and thus reducing the dimension. The downside is that the solution may change.

In the paper we defined a concept of safeness: Given an antimonotonic family $\iset{F}$ of itemsets a set $C$ of attributes is safe if the projection to $C$ does not change the solution of a query regardless of the query function and the given frequencies for $\iset{F}$. We characterised this concept by using graph theory. We also provided an efficient algorithm for finding the minimal safe set containing some given set.

We should point out that while our examples and experiments were focused on conjunctive queries, our theorems work with a query function of any shape

If the family of itemsets satisfies certain requirements, that is, it is triangulated and clique-safe, then we can obtain safe sets from junction trees. We also show that the factorisation obtained from a junction tree can be used to reduce the computational burden of the optimisation problem.

In addition, we provided a heuristic algorithm for finding restricted safe sets. The algorithm tries to construct a set of items such that the optimisation problem does not change for some \emph{given} itemset frequencies.

We ask ourselves: In practice, should we use the safe sets rather than the trivial projections? The advantage is that the (restricted) safe sets always produce outcome at least as good as the trivial approach. The downside is the additional computational burden. Our tests indicate that if a user makes a random query then in abt. $93\%-99\%$ of the cases the bounds are equal in both approaches. However, this comparison is unfair because there is a large number of queries where the projection sets are equal. To get the better picture we divide the queries into two classes \textsc{Trivial} and \textsc{Complex}, the first containing the queries such that the projections sets are equal, and the second containing the remaining queries. In the first class there is no improvement in the outcome \emph{but} there is no additional computational burden (checking that the set is safe is cheap comparing to the linear programming). If a query was in \textsc{Complex}, then in $10\%$ of the cases projecting on restricted safe sets did produce more tight bounds.
\label{sec:concl}
\begin{acknowledgements}
The author wishes to thank Heikki Mannila and Jouni Sep\-p\"anen for their helpful comments.
\end{acknowledgements}
\bibliography{bibliography}
\appendix
\section{Appendix}
This section contains the proofs for the theorems presented in the paper.
\subsection{Proof of Theorem~\ref{thr:suf}}
Let $\theta$ be any consistent frequencies for $\iset{F}$. Let $\iset{H} = \proj{C}{\iset{F}}$. To prove the theorem we will show that any distribution defined on items $C$ and satisfying the frequencies $\proj{C}{\theta}$ can be extended to a distribution defined on the set $A$ and satisfying the frequencies $\theta$.

Let $W = A - C$. Partition $W$ into connected blocks $W_i$ such that $x, y \in W_i$ if and only if there is a path $P$ from $x$ to $y$ such that $P \cap C = \emptyset$. Note that the items coming from the same $W_i$ have the same frontier. Therefore, $\front{W_i}{C}$ is well-defined. We denote $\front{W_i}{C}$ by $V_i$.

Let $p^{ME}$ be the maximum entropy distribution defined on the items $A$ and satisfying $\theta$. Note that there is no chord containing elements from $W_i$ and from $C-V_i$ at the same time. This implies that we can write $p^{ME}$ as
\[
p^{ME}(\rb{A}) = p^{ME}(\rb{C}) \prod_i \frac{p^{ME}\pr{\rb{W_i}, \rb{V_i}}}{p^{ME}\pr{\rb{V_i}}}.
\]
Let $p$ be any distribution defined on $C$ and satisfying the frequencies $\proj{C}{\theta}$. Note that $p^{ME}\pr{\rb{V_i}} = p\pr{\rb{V_i}}$, and hence we can extend $p$ to the set $A$ by defining
\[
p(\rb{A}) = p(\rb{C}) \prod_i \frac{p^{ME}\pr{\rb{W_i}, \rb{V_i}}}{p^{ME}\pr{\rb{V_i}}}.
\]
To complete the proof we will need to prove that $p$ satisfies the frequencies $\theta$. Select any itemset $U \in \iset{F}$. There are two possible cases: Either $U \subseteq C$, which implies that $U \in \iset{H}$ and since $p$ satisfies $\proj{C}{\theta}$ it follows that $p$ also satisfies $\theta_U$.

The other case is that $U$ has elements outside $C$. Note that $U$ can have elements in only one $W_i$, say, $W_j$. This in turn implies that $U$ cannot have elements in $C - \front{W_j}{C}$, that is, $U \subseteq W_j \cup V_i$. Note that $p^{ME}\pr{\rb{W_i}, \rb{V_i}} = p\pr{\rb{W_i}, \rb{V_i}}$. Since $p^{ME}$ satisfies $\theta$, $p$ satisfies $\theta_U$. This completes the theorem.

\subsection{Proof of Theorem~\ref{thr:nec}}
Assume that we are given a family $\iset{F}$ of itemsets and a set $C$ such that there exists $x \notin C$ such that $\front{x}{C} \notin \iset{F}$. Select $Y \subseteq \front{x}{C}$ to be some subset of the frontier such that $Y \notin \iset{F}$ and each proper subset of $Y$ is contained in $\iset{F}$. We can also assume that paths from $x$ to $Y$ are of length $1$. This is done by setting the intermediate attributes lying on the paths to be equivalent with $x$. We can also set the rest of the attributes to be equivalent with $0$. Therefore, we can redefine $C = Y$, the underlying set of attributes to consist only of $Y$ and $x$, and $\iset{F}$ to be
\[
\iset{F} = \set{Z \mid Z \subset C, Z \neq C} \cup \set{yx \mid y \in C}.
\]
Let $\theta = \set{\theta_Z \mid Z \in \iset{F}}$ be the frequencies for the itemset family $\iset{F}$ such that
\begin{equation}
\begin{array}{rcll}
\theta_Z & = & 0.5^{-\abs{Z}} &\text{if } Z \subset C \\
\theta_Z & = & 0.5 & \text{if } Z = x\\
\theta_Z & = & c &\text{if } Z = xy \text{ for } y \in C,
\label{eq:freqcond}
\end{array}
\end{equation}
where $c$ is a constant (to be determined later).

Define $n$ to be the number of elements in $C$. Let $k$ be the number of ones in the random bit vector $\rb{C}$. Let us now consider the following three distributions defined on $C$:
\begin{equation*}
\begin{array}{rcl}
p_1(\rb{C}) &=&
\left\{
\begin{array}{lll}
2^{-n+1} &, & n - k \text{ is even} \\
0 &, & n - k \text{ is odd}
\end{array}\right. \\
p_2(\rb{C}) &=& 2^{-n} \\
p_3(\rb{C}) &=&
\left\{
\begin{array}{lll}
2^{-n+1} &, & n - k \text{ is odd} \\
0 &, & n - k \text{ is even}
\end{array}\right.. \\
\end{array}
\end{equation*}
Note that all three distributions satisfy the first condition in Eq.~\ref{eq:freqcond}. Note also that $p_i(\rb{C})$ depends only on the number of ones in $\rb{C}$. We will slightly abuse the notation and denote $p_i(k) = p_i(\rb{C})$, where $\rb{C}$ is a random vector having $k$ ones.

Assume that we have extended $p_i(\rb{C})$ to $p_i(\rb{C}, \rb{x})$ satisfying $\theta$. We can assume that $p_i(\rb{C}, \rb{x})$ depends only on the number of ones in $\rb{C}$ and the value of $\rb{x}$. Define $c_i(n, k) = p_i(\rb{C}, \rb{x} = 1)$, where $\rb{C}$ is a random vector having $k$ ones. Note that
\[
0.5 = p_i(\rb{x} = 1) = \sum_{k=0}^n {n \choose k} c_i(n, k).
\]
If we select any attribute $z \in C$, then
\[
c = p_i(\rb{z}=1, \rb{x}=1) = \sum_{k=1}^n {n-1 \choose k-1} c_i(n, k).
\]
If we now consider the conditions given in Eq.~\ref{eq:freqcond} and require that $p_i(\rb{x}=1) = \theta_x = 0.5$ and also require that $p_i(\rb{z}=1, \rb{x}=1) = c$ is the largest possible, then we get the following three optimisation problems:
\begin{equation}
\begin{array}{lrcl}
\textsc{Problem P$_i$}: \\
\text{Maximise} & c_i(n) & = & \sum_{k=1}^n {n-1 \choose k-1} c_i(n, k) \\
\text{subject to} & c_i(n, k) & \geq & 0 \\
& c_i(n, k) & \leq & p_i(k) \\
& 0.5 & = & \sum_{k=0}^n {n \choose k} c_i(n, k)
\label{eq:problem}
\end{array}
\end{equation}
If we can show that the statement 
\[
c_1(n) = c_2(n) = c_3(n)
\]
is false, then by setting $c = \max(c_1(n), c_2(n), c_3(n))$ in Eq.~\ref{eq:freqcond} we obtain such frequencies that at least one of the distributions $p_i$ cannot be extended to $x$. We will prove our claim by assuming otherwise and showing that the assumption leads to a contradiction.

Note that ${n-1 \choose k-1}/{n \choose k} = k/n$. This implies that the maximal solution $c_2(n)$ has the \emph{unique} form
\begin{equation}
c_2(n, k) = \left\{\begin{array}{lll}
2^{-n} &, & k > \frac{n}{2} \\
2^{-n-1} &, & k = \frac{n}{2} \text{ and }  n \text{ is even} \\
0 &, & \text{otherwise.}
\end{array}
\right.
\label{eq:form2}
\end{equation}
Define series $b(n, k) = \frac{1}{2}\pr{c_1(n, k) + c_3(n, k)}$. Note that $b(n,k)$ is a feasible solution for Problem P$_2$ in Eq.~\ref{eq:problem}. Moreover, since we assume that $c_2(n) = c_1(n) = c_3(n)$, it follows that $b(n, k)$ produces the optimal solution $c_2(n)$. Therefore, $b(n, k) = c_2(n, k)$. This implies that $c_1(n, k)$ and $c_3(n, k)$ have the forms
\begin{equation}
\label{eq:form1}
c_1(n, k)  =  \left\{\begin{array}{lll}
2c_2(n, k) &, & n-k \text{ is even} \\
0 &, & n-k \text{ is odd} \\
\end{array}\right.
\end{equation}
\begin{equation}
\label{eq:form3}
c_3(n, k)  =  \left\{\begin{array}{lll}
2c_2(n, k) &, & n-k \text{ is odd} \\
0 &, & n-k \text{ is even} \\
\end{array}\right..
\end{equation}
Assume now that $n$ is odd. The conditions of Problems P$_1$ and P$_3$ imply that
\[
\sum_{k=0}^n {n \choose k} c_1(n, k) = 0.5 = \sum_{k=0}^n {n \choose k} c_3(n, k).
\]
By applying Eqs.~\ref{eq:form2}--~\ref{eq:form3} to this equation we obtain, depending on $n$, either the identity
\[
{n \choose n}+{n \choose n-2}+\ldots+{n \choose \frac{n+1}{2}} = {n \choose n-1}+{n \choose n-3}+\ldots+{n \choose \frac{n+3}{2}}
\]
or
\[
{n \choose n}+{n \choose n-2}+\ldots+{n \choose \frac{n+3}{2}} = {n \choose n-1}+{n \choose n-3}+\ldots+{n \choose \frac{n+1}{2}}.
\]
Both of these identities are false since the series having the term ${n \choose \frac{n+1}{2}}$ is always larger. This proves our claim for the cases where $n$ is odd.

Assume now that $n$ is even. The assumption $c_1(n) = c_3(n)$ together with Eqs.~\ref{eq:form2}--~\ref{eq:form3} implies the identity
\[
{n-1 \choose n-1}+{n-1 \choose n-3}+\ldots+\frac{1}{2}{n-1 \choose \frac{n}{2}-1} = {n-1 \choose n-2}+{n-1 \choose n-4}+\ldots+{n-1 \choose \frac{n}{2}}.
\]
We apply the identity 
\begin{equation}
\label{eq:pascal}
{n \choose k} = {n-1 \choose k} + {n-1 \choose k-1}
\end{equation}
to this equation and cancel out the equal terms from both sides. This gives us the identity
\[
\frac{1}{2}{n-1 \choose \frac{n}{2}-1} = {n-2 \choose \frac{n}{2}-1}.
\]
By applying again Eq.~\ref{eq:pascal} we obtain
\[
{n-2 \choose \frac{n}{2}-2} = {n-2 \choose \frac{n}{2}-1}.
\]
This is true for no $n$ and thus we have proved our claim.

\subsection{Proof of Theorem~\ref{thr:me}}
Denote by $\ent{p}$ the entropy of a distribution $p$. We know that $\ent{q^{ME}} \geq \ent{\proj{C}{p^{ME}}}$. Assume now that $q$ is a distribution satisfying the frequencies $\proj{C}{\theta}$. Let us extend $q$ as we did in the proof of Theorem~\ref{thr:suf}:
\[
p(\rb{A}) = q(\rb{C}) \prod_i \frac{p^{ME}\pr{\rb{W_i}, \rb{V_i}}}{p^{ME}\pr{\rb{V_i}}}.
\]
The entropy of this distribution is of the form $\ent{p} = \ent{q}+c$,
where 
\[
c = \sum_i \ent{\proj{W_i \cup V_i}{p^{ME}}} - \ent{\proj{V_i}{p^{ME}}}
\]
is a constant not depending on $q$. This characterisation is valid because $\proj{V_i}{p^{ME}} = \proj{V_i}{q}$.
If we let $q = q^{ME}$, it follows that
\[
\ent{p^{ME}} \geq \ent{p} = \ent{q^{ME}}+c \geq \ent{\proj{C}{p^{ME}}}+c.
\]
If we now let $q = \proj{C}{p^{ME}}$, it follows that $p = p^{ME}$ and this implies that $\ent{p^{ME}} = \ent{\proj{C}{p^{ME}}}+c$. Thus $\ent{q^{ME}} = \ent{\proj{C}{p^{ME}}}$. The distribution maximising entropy is unique, thus $\proj{C}{p^{ME}} = q^{ME}$.

\subsection{Proof of Theorem~\ref{thr:optimal}}
Assume that there is $x \in Z$ such that $x \notin Y$. Let $U_x = \enset{u_1}{u_L}$ be as it is defined in Algorithm~\ref{cd:search}. Let $P_i$ be the shortest path from $x$ to $u_i$ and define $v_i$ to be the first item on $P_i$ belonging to $Y$. There are two possible cases: Either $v_i = u_i$ which implies that $u_i \in \front{x}{Y}$, or $u_i$ is blocked by some other element in $Y$. If $U_x \subseteq \front{x}{Y}$, then the safeness condition is violated. Therefore, there exists $u_j$ such that $v_j \neq u_j$.

We will prove that $v_j$ outranks $x$, that is, $\rank{v_j}{C} > \rank{x}{C}$. It is easy to see that it is sufficient to prove that $\rank{v_j}{U_x} > \rank{x}{U_x}$. In order to do this note that $\enset{v_1}{v_L} \subseteq \front{x}{Y} \in \iset{F}$. Therefore, because of the antimonotonic property of $\iset{F}$, there is an edge from $v_j$ to each $v_i$. This implies that there is a path $R_i$ from $v_j$ to $u_i$ such that $\abs{R_i} \leq \abs{P_i}$, that is, the length of $R_i$ is smaller or equal than the length of $P_i$. Also note, that since $v_j$ lies on $P_j$, there exists a path $R_j$ from $v_j$ to $u_j$ such that $\abs{R_j} < \abs{P_j}$. This implies that $\rank{v_j}{U_x} > \rank{x}{U_x}$.

Also, note that $U_x \subset \nbhd{v_j}{r}$, where $r$ is the search radius defined in Algorithm~\ref{cd:search}. This implies that $v_j$ is discovered during the search phase, that is, $v_j$ is one of the violating nodes.

To complete the proof we need to show that $v_j$ is a neighbour of $C$. Since $x$ is a neighbour of $C$, there is $u_k$ such that there is an edge between $x$ and $u_k$. This implies that $v_k = u_k$. Since there is an edge between $v_j$ and $v_k$, it follows that $v_j$ is neighbour of $C$.

\subsection{Proof of Theorem~\ref{thr:junction1}}
Let $a$ be some item belonging to some inner clique $Q$ but not belonging in any inner separator. The clique $Q$ is unique and the only reachable items of $C$ from $a$ are the inner separators incident to $Q$. Since $Q$ is a clique, it follows from the clique-safeness assumption that the frontier of $a$ is included in $\iset{F}$.

Let now $a$ be any item that is not included in any inner clique. There exists a unique inner clique $Q$ such that all the paths from $a$ to $C$ go through this clique. This implies that the frontier of $a$ is again the inner separators incident to $Q$.

\subsection{Proof of Theorem~\ref{thr:junction3}}
\begin{figure}[ht!]
\center
\begin{minipage}{4cm}
\includegraphics[width=3cm]{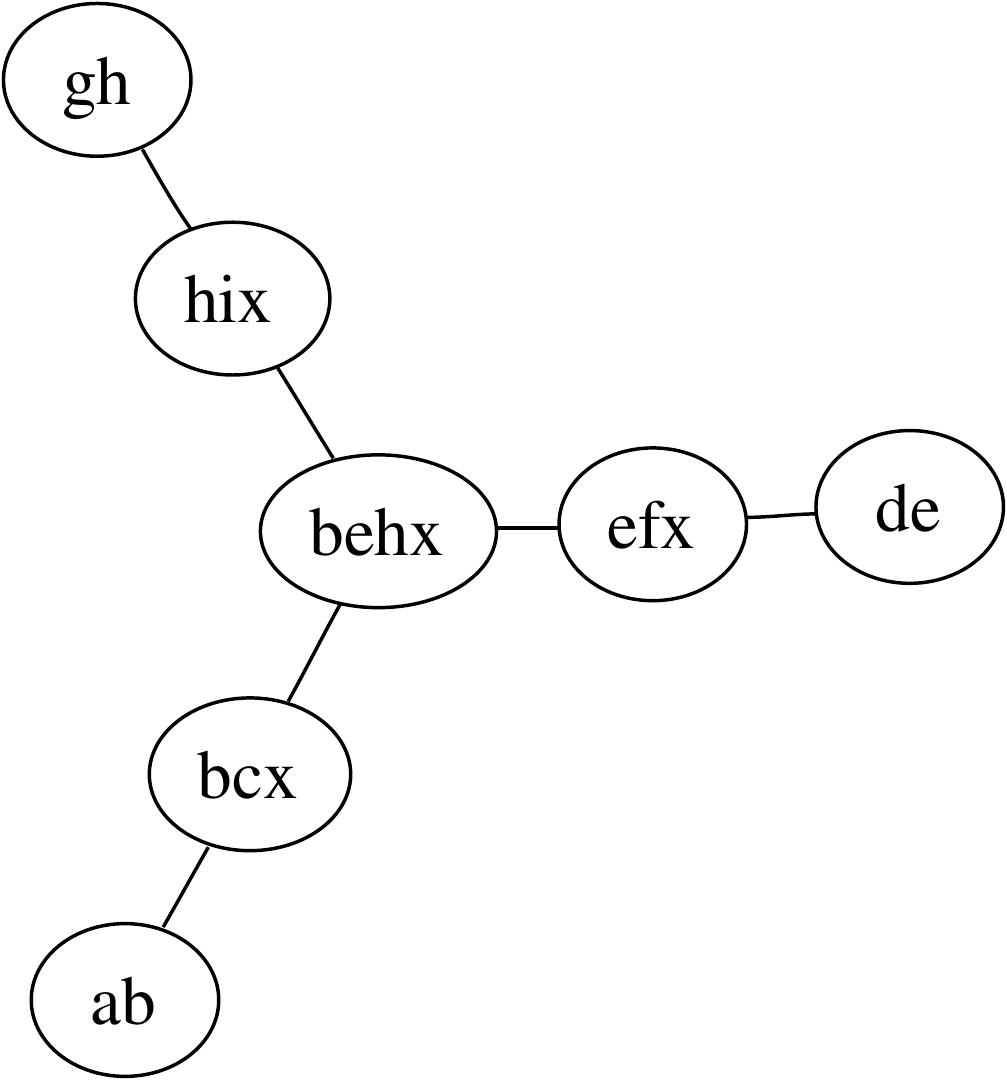}
\end{minipage}
\begin{minipage}{3cm}
\includegraphics[width=2.5cm]{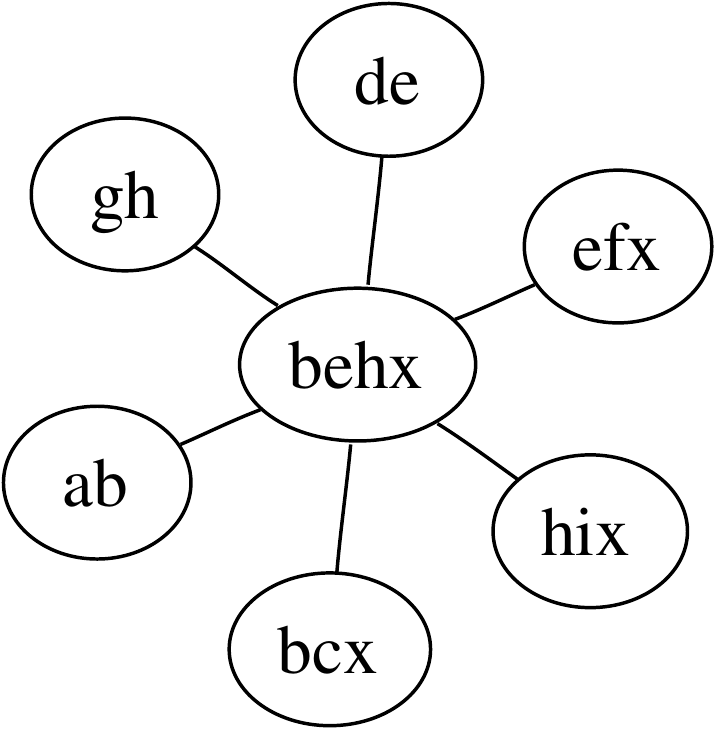}
\end{minipage}
\caption{Two equivalent junction trees. Our goal is to find the minimal safe set for $B = \set{a,d,g}$. The left junction tree is before the modification and the right is after the modification. We see that the attribute $x$ is not included in the inner separators in the right tree. The sets appearing in the proof are as follows: The minimal safe set $C$ is $adgbeh$. $I$ consists of $3$ separators $bx$, $ex$, and $hx$. The other separators belong to $J$. $V$ consists of $4$ cliques $bcx$, $efx$, $hix$, and $behx$. The clique $Q$ is $behx$.}
\label{fig:modify}
\end{figure}

We will prove that if we have an item $a$ coming from some inner separator and not included in the minimal safe set, then we can alter the junction tree such that the item $a$ is no longer included in the inner separators. For the sake of clarity, we illustrate an example of the modification process in Figure~\ref{fig:modify}.

Let $G$ be the dependency graph and $T$ the current junction tree. Let $C$ be the minimal safe set containing $B$ and let $a \notin C$ be an item coming from some inner separator. Let us consider paths (in $G$) from $a$ to its frontier. For the sake of clarity, we prove only the case where the paths from $a$ to $C$ are of length $1$. The proof for the general case is similar.

Let $I$ be the collection of inner separators containing $a$. Let $V$ be the collection of (inner) cliques incident to the inner separators included in $I$. The pair $(V, I)$ defines a subtree of $T$. Let $J$ be the set of inner separators incident to some clique in $V$ but not included in $I$. Note that each item coming from the inner separators included in $J$ must be included in $C$ because otherwise we have violated the assumption that the paths from $a$ to its frontier are of length $1$.

The frontier of $a$ consists of the items of the inner separators in $J$ and of possibly some items from the set $B$. By the assumption the frontier is in $\iset{F}$ and thus it is fully connected. It follows that there is a clique $Q$ containing the frontier. If $Q \notin V$, a clique from $V$ closest to $Q$ also contains the frontier. Hence we can assume $Q \in V$.

Select a separator $E \in J$. Let $U \notin V$ be the clique incident to $E$. We modify the tree by cutting the edge $E$ and reattaching $U$ to $Q$. The procedure is performed to each separator in $J$. The obtained tree satisfies the running intersection property since $Q$ contains the items coming from each inner separators included in $J$. If the frontier contained any items included in $B$, then $Q$ contains these items. It is easy to see that each clique in $V$, except for the clique $Q$, becomes outer. Therefore, $a$ is no longer included in any inner separator.

\subsection{Proof of Theorem~\ref{thr:fast}}
Let $\hat{p}$ be the optimal distribution. Then by marginalising we can obtain $\hat{p}_i$, and $\hat{q}_j$ which produce the same solution for the reduced problem.

To prove the other direction let $\hat{p}_i$, and $\hat{q}_j$ be the optimal distributions for the reduced problem. Since the running intersection property holds, we can define the joint distribution $\hat{p}$ by $\hat{p} = \prod_{i}\hat{p}_i / \prod_j \hat{q}_j$. It is straightforward to see that $\hat{p}$ satisfies the frequencies. This proves the statement.
\end{document}